\documentstyle[amssymb,preprint,aps,12pt]{revtex}
\newtheorem{theorem}{Theorem}

\newtheorem{lemma}[theorem]{Lemma}

\newtheorem{proposition}[theorem]{Proposition}

\newtheorem{solution}[theorem]{Solution}

\begin{document}
\title{Transfer matrices for scalar fields on curved spaces}
\author{E. Prodan}
\address{University of Houston, 4800 Calhoun Rd, Houston TX 77204-5508, USA}
\date{\today}
\maketitle

\begin{abstract}
We apply Nelson's technique of constructing Euclidean fields to the case of
classical scalar fields on curved spaces. It is shown how to construct a
transfer matrix and, for a class of metrics, the basic spectral properties
of its generator are investigated. An application concerning decoupling of
non-convex disjoint region is given.
\end{abstract}

\section{Introduction}

\noindent We start our construction from the ideas comprised in Nelson$%
\prime $s axioms \cite{Ne} for scalar Euclidean-Markoff quantum fields.
Here, the Markoff property of certain projectors is one of the basic
ingredient in defining the transfer matrix of whom generator is identified
with the Hamiltonian of Wightman quantum scalar field. We found that these
ideas can be used in the same way at the non-quantum level. In the case of
the scalar fields on Riemannian manifolds, for an arbitrary direction, we
construct a propagator by using the Markoff property. In the stationary case
it becomes a semigroup which can be considered as the transfer matrix of the
system and, further, it can be used in introducing a Hamiltonian. We will
show that the propagator is exponentially bounded by using Agmon$\prime $s 
\cite{Ag} results in exponential decay of solutions of second-order elliptic
equations. An application concerning the decoupling (in the sense of \cite
{Si1}) of two disjoint non-convex regions is given.

\section{Introductory definitions and results}

\noindent Let us consider the Riemannian manifold $\left( R^{n+1},g\right) $
and the Laplace-Beltrami operator on it, $\Delta $. For a point in $R^{n+1}$
we use the notation $\left( t,x\right) $. Let $E_{m}\left( t,x;s,y\right) $
be the kernel of $\left( \Delta +m^{2}\right) ^{-1}$ on $L^{2}\left( R^{n+1},%
\sqrt{g}dtdx\right) $. As in \cite{Di}, we will not consider the additional
term $%
%TCIMACRO{\dfrac{1}{6}}%
%BeginExpansion
{\displaystyle{1 \over 6}}%
%EndExpansion
\rho .$ One defines the space $N\subset {\cal D}^{\prime }\left(
R^{n+1}\right) $, $f\in N$ if : 
\begin{equation}
\left\| f\right\| _{N}^{2}=%
%TCIMACRO{\dint}%
%BeginExpansion
\displaystyle\int %
%EndExpansion
_{R^{n+1}}%
%TCIMACRO{\dint}%
%BeginExpansion
\displaystyle\int %
%EndExpansion
_{R^{n+1}}\bar{f}\left( t,x\right) E_{m}\left( t,x;s,y\right) f\left(
s,y\right) \sqrt{g\left( t,x\right) }\sqrt{g\left( s,y\right) }%
dtdxdsdy<\infty \text{,}
\end{equation}
and, for each $\sigma \in R$, let $N_{\sigma }\subset D^{\prime }\left(
R^{n}\right) $ be the space: $g\in N_{\sigma }$ if 
\begin{equation}
\left\| g\right\| _{N_{\sigma }}^{2}=%
%TCIMACRO{\dint}%
%BeginExpansion
\displaystyle\int %
%EndExpansion
_{R^{n}}\bar{g}\left( x\right) E_{m}\left( \sigma ,x;\sigma ,y\right)
g\left( y\right) \sqrt{g\left( \sigma ,x\right) }\sqrt{g\left( \sigma
,y\right) }dxdy<\infty \text{.}
\end{equation}
We will consider that, as in the Euclidean case, the space $L^{2}\left(
R^{n},d\mu _{\sigma }\right) \subset N_{\sigma }$, where $d\mu _{\sigma
}\left( x\right) =\sqrt{g\left( \sigma ,x\right) }d^{n}x$ and that it is
dense in $N_{\sigma }$ for each $\sigma \in R$. Now, let $\hat{E}_{\sigma
}:N_{\sigma }\rightarrow L^{2}\left( R^{n},d\mu _{\sigma }\right) $ be the
operator corresponding to the kernel $E_{m}\left( \sigma ,x;\sigma ,y\right) 
$. Then $\hat{E}_{\sigma }^{1/2}$ defines an isometry from $N_{\sigma }$ to $%
L^{2}\left( R^{n},d\mu _{\sigma }\right) $ and let $\left( \hat{E}_{\sigma
}^{1/2}\right) ^{\dagger }:L^{2}\left( R^{n},d\mu _{\sigma }\right)
\rightarrow $ $N_{\sigma }$ be its adjoint. The following are true: 
\begin{equation}
\hat{E}_{\sigma }^{1/2}\circ \left( \hat{E}_{\sigma }^{1/2}\right) ^{\dagger
}=1_{L^{2}\left( R^{n},d\mu _{\sigma }\right) }\text{ and }\left( \hat{E}%
_{\sigma }^{1/2}\right) ^{\dagger }\circ \hat{E}_{\sigma
}^{1/2}=1_{N_{\sigma }}\text{.}
\end{equation}
With our assumptions, $\hat{E}_{\sigma }^{1/2}\left( N_{\sigma }\right)
=L^{2}\left( R^{n},d\mu _{\sigma }\right) \subset N_{\sigma }$, the operator 
$\hat{E}_{\sigma }^{1/2}$ is bounded on $N_{\sigma }$. Moreover, one can
view $\left( \hat{E}_{\sigma }\right) ^{\dagger }$ as a dense defined
unbounded operator on $N_{\sigma }$, in fact, it is the inverse operator of $%
\hat{E}_{\sigma }$.

\noindent For $\sigma \in R$, let $j_{\sigma }$ be the operator $j_{\sigma
}:N_{\sigma }\rightarrow N$, $\left( j_{\sigma }\psi \right) \left(
t,x\right) =\psi \left( x\right) \delta \left( t-\sigma \right) $ and $%
j_{\sigma }^{\ast }$ be its adjoint. If $\Lambda $ is a closed subset of $%
R^{n+1}$ we denote by $N_{\Lambda }$ the subspace of $N$ which comprises all
distributions with support in $\Lambda $. The orthogonal projection of $N$
in $N_{\Lambda }$ will be denoted by $e_{\Lambda }$. Following \cite{Si2} we
have:

\begin{proposition}
The operators $j_{\sigma }$ are isometries and $j_{\sigma }^{*}j_{\sigma
}=1_{N_{\sigma }}$, $j_{\sigma }j_{\sigma }^{*}=e_{\sigma }$, where $%
e_{\sigma }$ denotes the projector corresponding to the subset of $R^{n+1}$, 
$t=\sigma $.
\end{proposition}

\noindent Then we define the operators: 
\begin{equation}
U_{\sigma ,\sigma ^{\prime }}:N_{\sigma ^{\prime }}\rightarrow N_{\sigma }%
\text{, }U_{\sigma ,\sigma ^{\prime }}=j_{\sigma }^{\ast }\circ j_{\sigma
^{\prime }}\text{.}  \label{!}
\end{equation}
We will derive in the following that $U_{\sigma ,\sigma ^{\prime }}$ are
propagators in the sense of \cite{Si3}. This will follow from the Markoff
property of the projectors $e_{\sigma }$.

\begin{lemma}
Let $A$, $B$ and $C$ be closed subsets in $R^{n+1}$ such that $C$ separates $%
A$ and $B$. Then $e_{A}\circ e_{C}\circ e_{B}=e_{A}\circ e_{B}$.
\end{lemma}

\begin{solution}
This is the consequence of the fact that $E_{m}$ is the kernel of a local
operator. The proof is identic with that of \cite{Si2}.
\end{solution}

\noindent The basics properties of $U_{\sigma ,\sigma ^{\prime }}$ operators
are stated in the following proposition.

\begin{proposition}
The family of operators $U_{\sigma ,\sigma ^{\prime }}$, $\sigma $, $\sigma
^{\prime }\in R$ has the following properties:\newline
1) $U_{\sigma ,\sigma ^{\prime }}\circ U_{\sigma ^{\prime },\sigma ^{\prime
\prime }}=U_{\sigma ,\sigma ^{\prime \prime }}$\newline
2) $U_{\sigma ,\sigma }=1_{N_{\sigma }}$\newline
3) $\left\| U_{\sigma ,\sigma ^{\prime }}\right\| \leqslant 1$.
\end{proposition}

\begin{solution}
1) Using the Markoff property we have: 
\begin{equation}
e_{\sigma }\circ e_{\sigma ^{\prime }}\circ e_{\sigma ^{\prime \prime
}}=e_{\sigma }e_{\sigma ^{\prime \prime }}\Leftrightarrow j_{\sigma }\circ
j_{\sigma }^{*}\circ j_{\sigma ^{\prime }}\circ j_{\sigma ^{\prime
}}^{*}\circ j_{\sigma ^{\prime \prime }}\circ j_{\sigma ^{\prime \prime
}}^{*}=j_{\sigma }\circ j_{\sigma }^{*}\circ j_{\sigma ^{\prime \prime
}}\circ j_{\sigma ^{\prime \prime }}^{*}.
\end{equation}
By composition with $j_{\sigma ^{\prime \prime }}$ at the right, we have 
\begin{equation}
j_{\sigma }\circ \left( j_{\sigma }^{*}\circ j_{\sigma ^{\prime }}\circ
j_{\sigma ^{\prime }}^{*}\circ j_{\sigma ^{\prime \prime }}-j_{\sigma
}^{*}\circ j_{\sigma ^{\prime \prime }}\right) =0.
\end{equation}
From the definition of $U_{\sigma ,\sigma ^{\prime }}$ and since $j_{\sigma
} $ are isometries, we conclude $U_{\sigma ,\sigma ^{\prime }}U_{\sigma
^{\prime },\sigma ^{\prime \prime }}=U_{\sigma ,\sigma ^{\prime \prime }}$.%
\newline
2) It follows from proposition 1.1 and definition of $U_{\sigma ,\sigma
^{\prime }}$.\newline
3) Because $j_{\sigma }^{*}$ and $j_{\sigma }$ are isometries, the property
results immediately.
\end{solution}

\section{Exponential bounds on propagators}

\noindent To improve our estimates on the propagators $U_{\sigma ,\sigma
^{\prime }}$ we need a supplementary condition on the metric $g$. We say
that an application $Q:R^{n+1}\rightarrow M\left( n+1,n+1\right) $ has
stable positivity if there exists $\varepsilon >0$ such that for any
application $\delta :R^{n+1}\rightarrow M\left( n+1,n+1\right) $ with $%
\left| \delta \left( x\right) ^{ij}\right| \leqslant \varepsilon $ the
matrices $Q\left( x\right) -\delta \left( x\right) $ are positive defined
for any $x\in R^{n+1}$. The following result is a direct application of
Agmon theory \cite{Ag} of exponentially decay of solutions of elliptic
second order operators.

\begin{proposition}
If the metric $g$ has stable positivity then for any $f\in N_{\sigma
^{\prime }}$: 
\begin{equation}
%TCIMACRO{\dint}%
%BeginExpansion
\displaystyle\int %
%EndExpansion
_{T_{0}}^{\infty }d\sigma \left\{ e^{\omega \sigma }\left\| \hat{E}_{\sigma
}^{1/2}\circ U_{\sigma ,\sigma ^{\prime }}f\right\| _{N_{\sigma }}\right\}
^{2}<\infty \text{,}
\end{equation}
provided $\omega <%
%TCIMACRO{\dfrac{m}{\sqrt{\sup g^{11}}}}%
%BeginExpansion
{\displaystyle{m \over \sqrt{\sup g^{11}}}}%
%EndExpansion
$.
\end{proposition}

\begin{solution}
Starting from 
\begin{equation}
\begin{array}{l}
\left\langle u,U_{\sigma ,\sigma ^{\prime }}f\right\rangle _{N_{\sigma
}}=\left\langle u,\hat{E}_{\sigma }\circ U_{\sigma ,\sigma ^{\prime
}}f\right\rangle _{L^{2}\left( R^{n},d\mu _{\sigma }\right) } \\ 
=\int_{R^{n}}\bar{u}\left( x\right) \left[ \int_{R^{n}}E_{m}\left( \sigma
,x;\sigma ^{\prime },y\right) f\left( y\right) d\mu _{\sigma ^{\prime
}}\left( y\right) \right] d\mu _{\sigma }\left( x\right)
\end{array}
\end{equation}
for $u\in N_{\sigma }$ and $f\in N_{\sigma ^{\prime }}$, it follows that $%
\varphi \left( \sigma ,x\right) =\left( \hat{E}_{\sigma }\circ U_{\sigma
,\sigma ^{\prime }}f\right) \left( x\right) $ is a solution of 
\begin{equation}
\left( \Delta +m^{2}\right) \varphi \left( \sigma ,x\right) =0
\end{equation}
for $\sigma >\sigma ^{\prime }$. Let $\rho _{m}\left( \cdot \,;\,\cdot
\right) $ denotes the distance corresponding to the metric $g_{m}=mg$. The
metric $g$ has stable positivity so, there is an $\varepsilon \in R_{+}$
such that $\rho _{m}\left( \sigma _{0},x_{0};\sigma ,x\right) >%
%TCIMACRO{\dfrac{\varepsilon }{m}}%
%BeginExpansion
{\displaystyle{\varepsilon  \over m}}%
%EndExpansion
\left| \sigma -\sigma _{0}\right| $. For $\Omega =\left\{ \left( \sigma
,x\right) :\sigma >T_{0}\right\} $, $T_{0}\in R_{+}$ and for some positive $%
\lambda $: 
\begin{equation}
\begin{array}{l}
\int_{\Omega }\left| \varphi \left( \sigma ,x\right) \right| ^{2}e^{-\lambda
\rho _{m}\left( T_{0},x_{0};\sigma ,x\right) }\sqrt{g\left( \sigma ,x\right) 
}d\sigma d^{n}x \\ 
=\int_{T_{0}}^{\infty }d\sigma \left\langle \hat{E}_{\sigma }\circ U_{\sigma
,\sigma ^{\prime }}f,\hat{E}_{\sigma }\circ U_{\sigma ,\sigma ^{\prime
}}f\right\rangle _{L^{2}\left( R^{n},d\mu _{\sigma }\right) }e^{-\lambda 
\frac{\varepsilon }{m}\left( \sigma -T_{0}\right) } \\ 
<ct.\int_{T_{0}}^{\infty }d\sigma \left\langle U_{\sigma ,\sigma ^{\prime
}}f,\hat{E}_{\sigma }\circ U_{\sigma ,\sigma ^{\prime }}f\right\rangle
_{L^{2}\left( R^{n},d\mu _{\sigma }\right) }e^{-\lambda \frac{\varepsilon }{m%
}\left( \sigma -T_{0}\right) } \\ 
=ct.\int_{T_{0}}^{\infty }d\sigma \left\| U_{\sigma ,\sigma ^{\prime
}}f\right\| _{N_{\sigma }}^{2}e^{-\lambda \frac{\varepsilon }{m}\left(
\sigma -T_{0}\right) }<\infty \text{.}
\end{array}
\end{equation}
So we are in the conditions of the main theorem of \cite{Ag}. It follows
that: 
\begin{equation}
\begin{array}{l}
\int_{\Omega }d\sigma d^{n}x\sqrt{g\left( \sigma ,x\right) }\left| \varphi
\left( \sigma ,x\right) \right| ^{2}\left( m^{2}-g\left( \nabla h\left(
\sigma ,x\right) ,\nabla h\left( \sigma ,x\right) \right) \right)
e^{2h\left( \sigma ,x\right) } \\ 
\leqslant \frac{2\left( 1+2d\right) }{d^{2}}m^{2}\int_{\Omega \setminus
\Omega _{d}}\left| \varphi \left( \sigma ,x\right) \right| ^{2}e^{2h\left(
\sigma ,x\right) }\sqrt{g\left( \sigma ,x\right) }dx\text{,}
\end{array}
\end{equation}
where $d$ is a positive number and $\Omega _{d}=\left\{ \left( \sigma
,x\right) \in \Omega :\rho _{m}\left( \left( \sigma ,x\right) ,\left\{
\infty \right\} \right) >d\right\} $. Here 
\begin{equation}
\rho _{m}\left( \left( \sigma ,x\right) ,\left\{ \infty \right\} \right)
=\sup \left\{ \rho _{m}\left( \left( \sigma ,x\right) ,\Omega \setminus
K\right) :K\text{ is a compact subset of }\Omega \right\} \text{.}
\end{equation}
The function $h$ is any function which satisfies the condition $g\left(
\nabla h\left( \sigma ,x\right) ,\nabla h\left( \sigma ,x\right) \right)
<m^{2}$. We choose $h\left( \sigma ,x\right) =\omega \sigma $ with $\omega <%
%TCIMACRO{\dfrac{m}{\sqrt{\sup g^{11}}}}%
%BeginExpansion
{\displaystyle{m \over \sqrt{\sup g^{11}}}}%
%EndExpansion
$. The above inequality becomes 
\begin{equation}
\begin{array}{l}
\int_{\Omega }d\sigma d^{n}x\sqrt{g\left( \sigma ,x\right) }\left| \varphi
\left( \sigma ,x\right) \right| ^{2}e^{2\omega \sigma } \\ 
<\frac{2\left( 1+2d\right) }{d^{2}}\frac{m^{2}}{m^{2}-\omega ^{2}}%
\int_{\Omega \setminus \Omega _{d}}d\sigma dx\sqrt{g\left( \sigma ,x\right) }%
\left| \varphi \left( \sigma ,x\right) \right| ^{2}e^{2\omega \sigma }\text{.%
}
\end{array}
\end{equation}
If for any point $\left( \sigma ,x\right) \in \Omega $ there is a geodesic
which starts in $\left( \sigma ,x\right) $ and ends in the hyperplane $%
\sigma =T_{0}$ then $\Omega \setminus \Omega _{d}\subset \left\{ \left( \tau
,x\right) :0<\sigma \leqslant T\right\} $ with $T$ sufficiently large but
finite. In conclusion 
\begin{equation}
\begin{array}{l}
\int_{\Omega }d\sigma d^{n}x\sqrt{g\left( \sigma ,x\right) }\left| \varphi
\left( \tau ,x\right) \right| ^{2}e^{2\omega \sigma } \\ 
=\int_{T_{0}}^{\infty }d\sigma e^{2\omega \sigma }\left\langle \hat{E}%
_{\sigma }\circ U_{\sigma ,\sigma ^{\prime }}f,\hat{E}_{\sigma }\circ
U_{\sigma ,\sigma ^{\prime }}f\right\rangle _{L^{2}\left( R^{n},\mu _{\sigma
}\right) }<\infty \text{,}
\end{array}
\end{equation}
or 
\begin{equation}
%TCIMACRO{\dint}%
%BeginExpansion
\displaystyle\int %
%EndExpansion
_{T_{0}}^{\infty }d\sigma e^{2\omega \sigma }\left\langle \hat{E}_{\sigma
}\circ U_{\sigma ,\sigma ^{\prime }}f,\hat{E}_{\sigma }\circ U_{\sigma
,\sigma ^{\prime }}f\right\rangle _{L^{2}\left( R^{n},\mu _{\sigma }\right)
}<\infty \text{,}
\end{equation}
which implies 
\begin{equation}
%TCIMACRO{\dint}%
%BeginExpansion
\displaystyle\int %
%EndExpansion
_{T_{0}}^{\infty }d\sigma \left\{ e^{\omega \sigma }\left\| \hat{E}_{\sigma
}^{1/2}\circ U_{\sigma ,\sigma ^{\prime }}f\right\| _{N_{\sigma }}\right\}
^{2}<\infty \text{.}
\end{equation}
\end{solution}

\section{The stationary case}

\noindent We consider in this section that there is a coordinate system such
that the metric $g$ is independent of first coordinate. In this case, the
spaces $N_{\sigma }$ and the operators $\hat{E}_{\sigma }^{1/2}$ are
identically and will be denoted by $N_{0}$ and $\hat{E}_{0}^{1/2}$
respectively. Thus, the operators $U_{\sigma ,\sigma ^{\prime }}$ are
defined on the same Hilbert space and depend only on the difference $\sigma
-\sigma ^{\prime }:$ $U_{\sigma ,\sigma ^{\prime }}=U_{\sigma -\sigma
^{\prime }}$. The family of operators $\left\{ U_{\tau }\right\} _{\tau \in
R_{+}}$ forms a semigroup. Using the results about existence and properties
of the generators of semigroups \cite{X}, we can obtain bounds directly on
the transfer matrix $U_{\tau }$.

\begin{proposition}
The semigroup $\left\{ U_{\tau }\right\} _{\tau \in R_{+}}$ is exponentially
bounded: $\left\| U_{\tau }\right\| _{N_{0}}<e^{-\tau \omega }$ provided $%
\omega <%
%TCIMACRO{\dfrac{m}{\sqrt{sup\,g^{11}}}}%
%BeginExpansion
{\displaystyle{m \over \sqrt{sup\,g^{11}}}}%
%EndExpansion
$.
\end{proposition}

\begin{solution}
Because we have found estimates on $\hat{E}_{0}^{1/2}\circ U_{\tau }$, we
will consider the operators $\tilde{U}_{\tau }=\hat{E}_{0}^{1/2}\circ
U_{\tau }\circ \left( \hat{E}_{0}^{1/2}\right) ^{\dagger }$, well defined on 
$L^{2}\left( R^{n},d\mu _{0}\right) $. Using the fact that $L^{2}\left(
R^{n},d\mu _{0}\right) $ is dense in $N_{0}$ we can extend these operators
by continuity on the space $N_{0}$. In this way we have build the semigroup $%
\left\{ \tilde{U}_{\tau }\right\} _{\tau \in R_{+}}$ which satisfies the
estimates of the precedent section: 
\begin{equation}
%TCIMACRO{\dint}%
%BeginExpansion
\displaystyle\int %
%EndExpansion
_{T_{0}}^{\infty }d\tau \left\{ e^{\omega \tau }\left\| \tilde{U}_{\tau
}\right\| _{N_{0}}\right\} ^{2}<\infty \text{,}
\end{equation}
for some $T_{0}>0$. So $\left\{ \tilde{U}_{\tau }\right\} _{\tau \in R_{+}}$
is exponentially bounded and in consequence \cite{X}, if $\tilde{K}$ is its
generator ($\tilde{U}_{\tau }=e^{-\tau \tilde{K}}$) the resolvent set of $%
\tilde{K}$ satisfies: 
\begin{equation}
\left\{ z\in C\shortmid 
%TCIMACRO{\func{Re}}%
%BeginExpansion
\mathop{\rm Re}%
%EndExpansion
z\in (-\infty ,\omega )\right\} \subset \rho \left( \tilde{K}\right) \text{.}
\end{equation}
If $K$ is the generator of $\left\{ U_{\tau }\right\} _{\tau \in R_{+}}$
then, on ${\cal D}\left( K\right) $ we have: 
\begin{equation}
K=\left( \hat{E}_{0}^{1/2}\right) ^{\dagger }\circ \tilde{K}\circ \hat{E}%
_{0}^{1/2}
\end{equation}
by using the reciprocal formula 
\begin{equation}
U_{\tau }=\left( \hat{E}_{0}^{1/2}\right) ^{\dagger }\circ \tilde{U}_{\tau
}\circ \hat{E}_{0}^{1/2}\text{,}  \label{!!}
\end{equation}
valid on $N_{0}$. If the operator 
\begin{equation}
\left( \hat{E}_{0}^{1/2}\right) ^{\dagger }\circ \left( \tilde{K}-z\right)
^{-1}\circ \hat{E}_{0}^{1/2}  \label{!!!!}
\end{equation}
is well defined, even on a dense subset of $N_{0}$, then $K-z$ is
inversable. From 20 it follows that, if $\left( \tilde{K}-z\right) ^{-1}$
exists, then:
\begin{equation}
\left( \tilde{K}-z\right) ^{-1}\left( L^{2}\left( R^{n},d\mu _{0}\right)
\right) \subset L^{2}\left( R^{n},d\mu _{0}\right) \text{,}
\end{equation}
and in consequence $\left( \hat{E}_{0}^{1/2}\right) ^{\dagger }\circ \left( 
\tilde{K}-z\right) ^{-1}\circ \hat{E}_{0}^{1/2}$ is well defined on the
entire $N_{0}$. Will follow that $\rho \left( \tilde{K}\right) \subset \rho
\left( K\right) $ and this ends the proof.
\end{solution}

\noindent If the metric is symmetric at transformation $x^{1}\rightarrow
-x^{1}$, the transfer matrix generator is self-adjoint and it can be
considered as the Hamiltonian of the scalar field.

\section{Application}

\noindent Our application is for the Euclidean case. The results concerning
decoupling of different regions in quantum Euclidean fields are based
primarily on estimates of $\left\| e_{\Lambda _{1}}e_{\Lambda _{2}}\right\|
_{N}$, where $\Lambda _{1}$, $\Lambda _{2}$ are two disjoint regions. Let us
consider the two dimensional case. The most difficult case is when $\Lambda
_{1}$, $\Lambda _{2}$ are not convex and there is no possibility of drawing
a straight line between the two subsets. We can sharpen the existent
estimates \cite{Si2} for these cases by using the previous results. The idea
is to make a change of coordinates such that for the new coordinates, lines
like $\sigma =ct.$ separate the two sets and they are as closed as possible
to the boundaries of $\Lambda _{1}$, $\Lambda _{2}$. Then we can use the
exponential bounds of the previous section to evaluate $\left\| e_{\Lambda
_{1}}e_{\Lambda _{2}}\right\| _{N}$. More precisely:

\begin{proposition}
Let $\Lambda _{1}$, $\Lambda _{2}$ two regions in $R^{2}$ such that the
construction of the coordinates \ref{c} to be possible (after a rotation if
necessary). Then 
\begin{equation}
\left\| e_{\Lambda _{1}}\circ e_{\Lambda _{2}}\right\| _{N}\leqslant
e^{-m\left| \beta -\alpha \right| \min \left| \cos \theta \right| }\text{,}
\end{equation}
where $\theta $ and $\left| \beta -\alpha \right| $ will be defined during
the proof.
\end{proposition}

\begin{solution}
Let $\left( t,x\right) $ denotes the original coordinates in which the
metric is diagonal. Let $\gamma :R\rightarrow R^{2}$ be a curve which
separates $\Lambda _{1}$, $\Lambda _{2}$ and $\gamma \left( 0\right) =\left(
t=0,x=0\right) $. We define a new coordinate system $\left( \sigma ,\xi
\right) $ by 
\begin{equation}
\left\{ 
\begin{array}{c}
t\left( \sigma ,\xi \right) =\sigma +\gamma ^{1}\left( \xi \right) \\ 
x\left( \sigma ,\xi \right) =\gamma ^{2}\left( \xi \right)
\end{array}
\right.  \label{c}
\end{equation}
In the new coordinates, the metric is 
\begin{equation}
g^{\prime }\left( \sigma ,\xi \right) =\left( 
\begin{array}{cc}
1 & 
%TCIMACRO{\dfrac{d\gamma ^{1}}{d\xi } }%
%BeginExpansion
{\displaystyle{d\gamma ^{1} \over d\xi }}%
%EndExpansion
\\ 
%TCIMACRO{\dfrac{d\gamma ^{1}}{d\xi } }%
%BeginExpansion
{\displaystyle{d\gamma ^{1} \over d\xi }}%
%EndExpansion
& \left( 
%TCIMACRO{\dfrac{d\gamma ^{1}}{d\xi }}%
%BeginExpansion
{\displaystyle{d\gamma ^{1} \over d\xi }}%
%EndExpansion
\right) ^{2}+\left( 
%TCIMACRO{\dfrac{d\gamma ^{2}}{d\xi }}%
%BeginExpansion
{\displaystyle{d\gamma ^{2} \over d\xi }}%
%EndExpansion
\right) ^{2}
\end{array}
\right)
\end{equation}
so we are in the conditions of the last section. Using the Markoff property, 
\begin{equation}
\left\| e_{\Lambda _{1}}\circ e_{\Lambda _{2}}\right\| _{N}=\left\|
e_{\Lambda _{1}}\circ e_{\alpha }\circ e_{\beta }\circ e_{\Lambda
_{2}}\right\| _{N}\leqslant \left\| e_{\alpha }\circ e_{\beta }\right\| _{N}%
\text{,}
\end{equation}
where the lines $\sigma =\alpha $, $\sigma =\beta $ separate $\Lambda _{1}$
and $\Lambda _{2}$ exactly in the order they appear in the above relation
(in the sense that $\sigma =\alpha $ separates $\Lambda _{1}$ by $\sigma
=\beta $ etc.). Further 
\begin{equation}
\left\| j_{\alpha }\circ j_{\alpha }^{\dagger }\circ j_{\beta }\circ
j_{\beta }^{\dagger }\right\| _{N}=\left\| j_{\alpha }\circ U_{\alpha -\beta
}\circ j_{\beta }^{\dagger }\right\| _{N}=\left\| U_{\alpha -\beta }\right\|
_{N_{0}}.
\end{equation}
The element $\left( g^{\prime }\right) ^{11}$ is given by $\left( g^{\prime
}\right) ^{11}=%
%TCIMACRO{\dfrac{1}{\cos ^{2}\theta }}%
%BeginExpansion
{\displaystyle{1 \over \cos ^{2}\theta }}%
%EndExpansion
$, where $\theta $ is the angle between the tangent to the curve $\gamma $
and the $x$ axis. Using the bounds of the last section we have 
\begin{equation}
\left\| e_{\Lambda _{1}}\circ e_{\Lambda _{2}}\right\| _{N}\leqslant
e^{-m\left| \beta -\alpha \right| \min \left| \cos \theta \right| }.
\end{equation}
Performing first a rotation, one can choose the best values for $\left|
\beta -\alpha \right| $ and $\min \left| \cos \theta \right| $.
\end{solution}

\section{Conclusions}

\noindent Our primary goal was to define the transfer matrix for scalar
fields on curved spaces and to investigate the basic spectral properties of
its generator. Even though the generator is not self-adjoint in the general
case, this approach allows us to investigate this problem by using at least
two new tools besides the methods of Green functions. One is the
perturbations of hypercontractive semigroups \cite{Si4} and the other is the
adiabatic theorem.

\noindent Now it is straightforward to quantize the field by defining the
Markoff field over the space $N$. For the stationary, symmetric at time
reflection case (static), we think that one has now all elements to
construct the physical field (for example that proposed in \cite{Di}) by
following Nelson reconstruction method and holomorphic continuation of the
transfer matrix. Note that, acording to results of \cite{Si3}, the
holomorphic continuation of the transfer matrix to real time is still
possible, in the stationary case without symmetry at time reflection, as
long the spectrum of the generator belongs to the real axis. Of course, one
has to check that the results of \cite{Ne2} (sistematized in \cite{Si2}),
which are the core of the reconstruction theorem, are still valid. For the
general case, we think that the adiabatic theorem, especially the adiabatic
reduction theory \cite{Nen}, may play an important role in defining the
physical quantum field by following Nelson$\prime $s approach.


\begin{references}
\bibitem{Ne}  Nelson E., J. Func. Anal. {\bf 12}, 97 (1973)

\bibitem{Ag}  Agmon S., {\it Lectures on exponential decay of \ solutions of
second-order elliptic equations}, Princeton: Princeton Univ. Press (1982)

\bibitem{Si1}  Guerra F., Rosen L., Simon B., Ann. Math. {\bf 101}, 111
(1975)

\bibitem{Di}  Dimock J., J. Math. Phys. {\bf 20}, 2549 (1979)

\bibitem{Si2}  Simon B., {\it The} $P\left( \phi \right) _{2}$ {\it %
Euclidean (Quantum) Field Theory}, Princeton: Princeton University Press
(1974)

\bibitem{Si3}  Reed M., Simon B., {\it Methods of Modern Mathematical Physics%
}, Vol. 2, New York: Academic Press (1975)

\bibitem{X}  Neerven J., {\it The asymptotic behavior of semigroups of
linear operators}, Basel, Boston: Birkhauser (1996)

\bibitem{Si4}  Simon B., Hoegh-Krohn R., J. Func. Anal. {\bf 9}, 121 (1972)

\bibitem{Ne2}  Nelson E., J. Func.{\it \ }Anal. {\bf 11}, 211 (1972)

\bibitem{Nen}  Nenciu G., Commun. Math. Phys. {\bf 152}, 479 (1993)
\end{references}
\end{document}